# Exotics and PWA for πN Scattering

**Yakov Azimov**

*Petersburg Nuclear Physics Institute*
*Gatchina, Russia 188300*
*E-mail: azimov@thd.pnpi.spb.ru*

**Igor Strakovsky**

*The George Washington University*
*725 21$^{st}$ Street NW Washington, DC 20052 USA*
*E-mail: igor@gwu.edu*

Our talk is intended for the session in memory of Mitya Diakonov. The problem is considered of existence of flavor multiplets consisting of multi-quark baryons. We have argued that the S-matrix should have poles, at least Regge ones, with any quantum numbers, including exotic. This is a novel argument for possible reality of hadrons with arbitrary exotic structure. Though it does not provide a proof, yet there are no theoretical arguments to forbid exotics. Then we apply the partial-wave analysis (PWA) with addressing mainly to the non-strange exotic members of the anti-decuplet or even of higher multiplets. It suggested new N(1680) as a possible partner of $\Theta^+$. Later independent measurements of several collaborations seem to support our finding for N(1680), though its quantum numbers still wait for investigation. Similar approach to $\pi^+$-proton scattering, in combination with $K^+$-proton scattering, suggests a number of candidates for 27-plets. Their interesting feature is possible existence of both very wide and very narrow baryon states. This phenomenon may be analogous to recent Weinberg's suggestion for tetra-quark mesons. The situation is worth of further theoretical and experimental studies.

It would be an impossible task to discuss here all aspects of Mitya Diakonov's activity, so we will concentrate on the problem of multi-quark hadrons associated mainly with the *renowned* paper[1].

At present, we know 6 quark species (u, d, s, c, b, t). Every known baryon is still considered as a 3-quark system (q,q,q), while every known meson as a (q,anti-q)-system. Such a picture limits possibile quantum numbers. For instance, strangeness for baryons may be only S = 0, –1, –2, –3, while for mesons S = –1, 0, +1. However, a fixed number of constituents is in conflict with relativistic description (they should be appended by an arbitrary number of ``virtual'' particles). Additional q-anti-q pairs might provide hadrons with different (exotic) quantum numbers (for example, a baryon with S = +1). Let us see what Theory and Experiment say…

It is generally believed that strong interactions are ruled by Quantum ChromoDynamics (QCD) based on quarks and gluons. QCD seems to provide no restrictions on multi-quark exotic hadrons (at least, we do not know yet any restriction: *Why are there no strongly bound exotic states…, like those of two quarks and two anti-quarks or four quarks and one anti-quark?*[2]. Every QCD inspired model predicts such states, with properties strongly model-dependent. But experiments, for long time, had not given any solid evidence, and the searches have been stopped in 1986 or so.

In about that time, Diakonov, Petrov, and crew started to develop a new approach to strong interactions. They integrated out the gluon and light quark fields of QCD, replacing them by the meson field, while baryons, as Skirme suggested, correspond to solitons of this field. Such approach appears to work quite well for low lying baryonic states – octet and decuplet.

The soliton approach predicts a new multiplet, baryon anti-decuplet with relative small masses, including an exotic baryon with S = +1. Its mass uncertainty was too large, 1500 – 1600 MeV, and full width was expected ~100 MeV, as $\Delta(1232)$ has. Such predictions had not tempted experimentalists.

Diakonov, Petrov, and Polyakov[1] moved the task on. Using this quark-soliton approach and assuming the PDG baryon N(1710) to be a non-strange member of the anti-decuplet, they predicted $Z^+$ (or $\Theta^+$, according to Diakonov's suggestion of April 2003) with $J^P = \frac{1}{2}^+$, S = +1 to have M = 1530 MeV, and estimated its $\Gamma$ < 15 MeV. It was the first rather certain theoretical suggestion for parameters of this state. Experimentalists received a strong impetus to look for a narrow peak in $pK^0$ and/or $nK^+$ masses. Later Diakonov and Petrov made even stronger prediction[3], $\Gamma$ < 4 MeV. The SAID modified KN PWA suggested a longer phenomenological life time[4], $\Gamma \leq 1$ MeV, and DIANA reported the ``direct'' measurement[5], $\Gamma = 0.39 \pm 0.10$ MeV. Do we see a new kind of hadrons?

<u>Experiment</u>: There are more than ten papers with positive evidences and more than ten papers with negative results (some of them have high statistics and look more impressive). Present common opinion and that of PDG (since edition of 2008) is: *Pentaquarks are dead and absent*! (<u>Paradox</u>: There is, nevertheless, a great enthusiasm in searches for tetra-quarks! For instance, several JLab projects.)

More recent experimental results on the $\Theta^+$: CLAS does not see it (several papers on different reactions)[6]; LEPS[7] and DIANA[5,8] confirm their results. Comparative analysis: all the data, positive and negative, can be reconciled under a specific assumption on the production mechanism[9]. <u>Hypothesis</u>: Exotic (multiquark) hadrons are mainly produced through initial many-parton states (higher Fock components). Such states are always related to short-term fluctuations – in this sense, production of exotic hadrons may be viewed as a new kind of hard processes (in addition to DIS, Drell-Yan and so on). Then the $\Theta^+$ signal (if it exists at all) may be small indeed, and its amplification would be desirable.

To enhance a weak signal, one can use its interference with a stronger signal. Clear cases of interference are visible in reactions $e^+e^- \to$ hadrons[10]. Different resonances may interfere if their decays provide the same final states (direct interference, when all final particles can be decay products of any of the interfering resonances). Then resonance manifestations may be essentially distorted vs. the standard Breit-Wigner form. The same resonances may interfere differently in different decay modes. Such a kind of interference appears very efficient to search for rare decays of known resonances.

There exist other kinds of interference, where only some of final particles may come from any of the two interfering resonances. Nevertheless, different resonance configurations may produce the same final states of three or more particles, which are also coherent and may interfere (rearrangement interference)[10]. The interference result here depends on energies and momentum transfers; it may shift and move positions of bumps/dips. The phenomenon is known since 60's and was considered as hindrance to resonance studies. That is why such interference was usually (before and till now) cut away. Meanwhile, direct interference of resonances has become an efficient instrument actively used to study rare decays of known resonances. Amarian, Diakonov, and Polyakov have suggested to apply the rearrangement interference for revealing faint resonance signals (due to their enhancement by interference with a stronger signal)[11].

Reaction $\gamma p \to pK^0\bar{K}^0$ was studied by CLAS to look for $\Theta^+$ in the dedicated experiment[6]. Strong $\phi$–peak has been cut out ($M_X(p) > 1.04$ GeV), in accordance with tradition. No signal of $\Theta^+$ was seen, only hard restriction was given for its production cross section. In the new analysis[12], the same set of CLAS events is used to study the same reaction $\gamma p \to pK_S K_L$ <u>under</u> the $\phi$–peak: $M_X(p) = 1.02 \pm 0.01$ GeV. Really, the final state is $p\pi^+\pi^- X$. $K_S$ is reconstructed by the peak in the two-pion mass, and $K_L$ by peak in $M_X(pK_S)$. Signal in $M_X(K_S) = M(pK_L)$ is well seen, when applying more cuts (together or separately): for $M(pK_S)$ - to suppress kinematic reflections from known $\Sigma^{*}$'s (see PDG); for momentum transfer - to separate a definite $\phi$-production mechanism. MC for background uses the physical model for $\phi$-photoproduction at small momentum transfers[13], theoretically meaningful, experimentally adequate. Data reveal the peak at $M(pK_L) \sim 1.54$ GeV, its width consistent with the resolution. Significance of the peak is estimated to be 5.3$\sigma$. $M(pK_S)$ does not reveal an observable signal (worse resolution, in consistency with MC). If the signal is due to interference of the $\phi$-meson with some new state of $S = \pm 1$ (decay to $pK_L$), then it could be either $\Theta^+$ ($S = +1$), or an unknown $\Sigma^{*+}$ ($S = -1$) (would be extremely unusual – very narrow, all decay modes without kaons suppressed). It is thus seen that rearrangement interference may indeed amplify a faint signal of a resonance (with any quantum numbers! the faintness could be because of either small branching or suppressed production). For final confirmation of the pentaquark existence one needs to find also ist direct (not only interference) signal and determine its strangeness (the corresponding preliminary data are shown in a conference talk of LEPS).

What about a non-strange member of the anti-decuplet? Is it N(1710), as initially assumed[1], or some other? There are two critical points here: i) Are N(1710) and N(1680) at PDG12 Listings[14] the same or they are different? ii) N(1680), if exists, is the first narrow N*; how was it suggested? Conventional PWA (by construction) tends to miss resonances with $\Gamma < 30$ MeV[15]. To modify PWA, we first assume existence of a narrower resonance, add it to the amplitude, then re-fit over the whole database[15,16]. If refitting provides a worse description then resonance with corresponding M and $\Gamma$ is not supported. In the case of better description: i) Resonance may exist; ii) Effect can be due to various corrections (*e.g.*, thresholds); or iii) Both possibilities can contribute. Thus, some additional checks are necessary. One of them may be the fact that a true resonance should provide the effect only in a particular partial wave while non-resonance sources may show similar effects in various partial waves.

$\Delta\chi^2$ due to insertion of a πN resonance into $P_{11}$ ($J^P = 1/2^+$) drops down somewhere[16]. Two candidates are $M_R$ = 1680 MeV and 1730 MeV, $\Gamma_{\pi N}$ < 0.5 MeV and < 0.3 MeV, respectively. At $|M_R - W|$ >> $\Gamma_R$, the resonance contributes ~ $\Gamma_{el}/(M_R - W)$, so the procedure is less sensitive to $\Gamma_{tot}$ than to $\Gamma_{\pi N}$. No effects at M = 1680 MeV and possible (small) effect at M = 1730 MeV are seen for $S_{11}$ and $P_{13}$ waves. Expected decay properties of N(1680) are essentially model-dependent[16].

Several independent measurements agree with each others and support expectation: narrow structures are seen in γn→nη: GRAAL: W ≈ 1680 MeV, Γ < 30 MeV[17]; ELPH: W ≈ 1666 MeV, Γ < 40 MeV[18]; CB-ELSA: W ≈ 1685 MeV, Γ < 50 MeV[19]; and CB@MAMI: W ≈ 1675 MeV, Γ < 40 MeV[20]. All visible widths are consistent with experimental resolutions. There is only Fermi motion accounted for, without FSI applied. GRAAL studied Compton on neutron as well: W ≈ 1685 MeV, Γ < 30 MeV[21]. Some evidence in KΛ is also present[22]. New experiments and analyses are in progress.

<u>Intermediate Conclusion</u>: *The report of Theta's death was an exaggeration* (rephrasing of Mark Twain). Experimental evidences for the Θ-pentaquark existence seem to revive. One more anti-decuplet member, nucleon-like, seems to be revealed (in support to the $\Theta^+$). Investigation of multiquark hadrons may open new directions both for hadron spectroscopy and for QCD studies in general. The Θ-baryon and its companions, if further confirmed, may become a memorial to Mitya Diakonov.

Legacy of complex angular momenta allows one to show[23] that the S-matrix must have Regge poles with arbitrary flavor numbers, even exotic ones. This might hint at existence of exotic hadrons as well, being members of an arbitrary unitary multiplet. That is why we tried to apply the PWA approach to search also for higher multiplets, 27-plets[23]. All members of the same 27-plet should be ``correlated'': the same spin-parity and nearby masses. KN scattering with I=1 may see $\Theta_1$ from 27, while Δ from 27 or 10 may be related to scattering of πN with I = 3/2. Thus, a way to search for 27-plets: Look through scattering data and PWA for correlated pairs ($\Theta_1$, Δ), having the same spin-parity and nearby masses.

Published are one conventional PWA for KN[24] and several PWA's for πN scatterings. They give two pairs of correlated poles corresponding to broad resonances with heavier masses[23]. On the other side, the modified PWA's provided an unexpectedly large number of lighter candidates for narrow Δ- and $\Theta_1$-like states in πN and KN scattering[23]. The candidates are seen for any investigated spins and parities at several masses. Members of pairs (Δ, $\Theta_1$) have masses very close to each other and very small $\Gamma^{el}$, smaller than expected and even smaller than for $\Theta^+$(1540). In particular, we see several candidates for narrow $\Theta_1$ near 1530 MeV. Note that STAR gave preliminary evidence[25] for a $K^+p$ peak with M = 1530 MeV, $\Gamma_{tot}$ < 15 MeV, and status 4.2 σ.

<u>Overall</u>: Conventional and modified PWA's suggest many candidates for the ``correlated'' pairs (Δ, $\Theta_1$), each of which may label the corresponding 27-plet. There are candidates for both positive and negative parity multiplets, the latter not studied in details by soliton approaches (non-rotational excitations?). There appear two sets of candidates: very broad (and heavier), and very narrow (and lighter). Are there two kinds of decay dynamics? Similar problem may exist for tetra-quark mesons[26].

<u>Final Conclusion</u>: ``*...either these* [multiquark] *states will be found by experimentalists, or our confined, quark-gluon theory of hadrons is as yet lacking in some fundamental ingredient...*''[27] Rejection of the $\Theta^+$(1530) will retain open questions; on the other side, studies of the $\Theta^+$(1530), if confirmed (in Japan ?), will give important information both on hadron structure and on properties of QCD. Production of multiquark hadrons might be a new kind of hard processes, if it is related indeed with higher Fock components. Such hypothesis can suggest new interesting experiments.

This work was partly supported by the US DOE Grant No.DE-FG02-99ER41110 and Russian State Grant RSGSS-4801.2012.2.

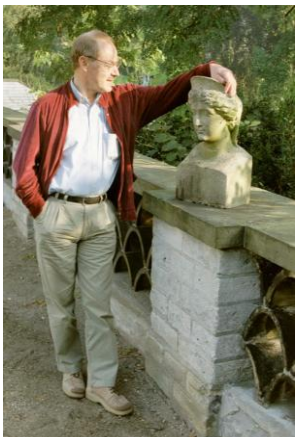 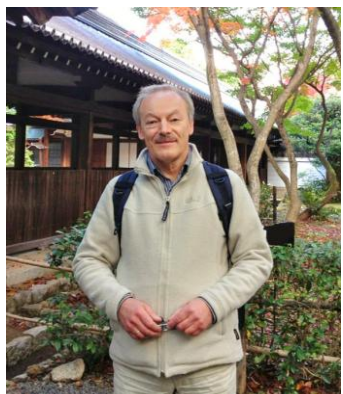 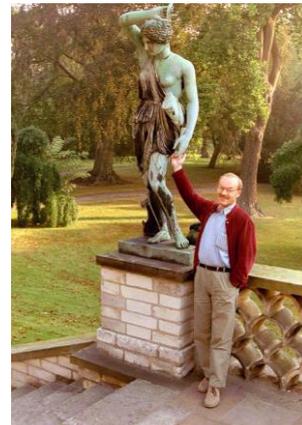

In Copenhagen, circa 2000.　　　In Kyoto, November 2012.　　　In Copenhagen, circa 2000.